\RequirePackage{fix-cm}

\documentclass[twocolumn]{svjour3}

\smartqed

\usepackage{graphicx}
\usepackage{natbib}   

\journalname{Cognitive Neurodynamics}

\begin{document}

\title{Network properties of healthy and Alzheimer's brains}

\author{Jos\'e C. P. Coninck$^1$ \and Fabiano A. S. Ferrari$^2$ \and Adriane S.
Reis$^3$ \and Kelly C. Iarosz$^4$ \and Antonio M. Batista$^5$ \and Ricardo L.
Viana$^3$}

\institute{$^1$Technolgical University of Paran\'a \\ Department of Statistics
\at Curitiba, Brazil 
\and $^2$Federal Univesity of the Valleys of Jequitinhonha and Mucuri,\\
Institute of Enginnering, Science and Technology \at Jana\'uba, Brazil
\and $^3$Federal University of Paran\'a, \\ Department of Physics \at Curitiba,
Brazil
\and $^4$University of S\~ao Paulo, \\ Institute of Physics \at S\~ao Paulo,
Brazil
\and $^5$State University of Ponta Grossa, \\ Department of Mathematics and
Statistics \at Ponta Grossa, Brazil}

\date{Received: date / Accepted: date}

\maketitle

\begin{abstract}
Small-world structures are often used to describe structural connections in the
brain. In this work, we compare the strucutural connection of cortical areas of
a healthy brain and a brain affected by Alzheimer's disease with artificial
small-world networks. Based on statistics analysis, we demonstrate that similar
small-world networks can be constructed using Newman-Watts procedure. The
network quantifiers of both structural matrices are identified inside the
probabilistic valley. Despite of similarities between strcutural connection
matrices and sampled small-world networks, increased assortivity can be found
in the Alzheimer brain. Our results indicate that network quantifiers can be
helpful to identify abnormalities in real structural connection matrices.
\end{abstract}


\section{Introduction}

One of the first fully reported neural network was the worm C. elegans
\citep{White1986}. The nervous system of C. elegans consists of 302 neurons
connected through $5000$ chemical and $600$ electrical synapses. Wa\-tts and
Strogatz showed that the C. elegans brain network can be described by a
small-world network \citep{Watts1998,Varshney2011}. Small-world networks are
characterized by high clustering and short average distance between nodes. They
have been observed in brain networks of animals and humans
\citep{Sporns2004,Bassett2006,Stam2014,Medina2008}. Evidences of small-world
properties can also be found in ensembles of neurons {\it in vitro}
\citep{Bettencourt2007}. The small worldness of neuronal networks is
hypothesized to be a consequence of optimization process associated with
minimal wiring cost, robustness and balance between local processing and global
integration \citep{Reijneveld2007,Bullmore2012}. 

Brain networks can be obtained in different levels, such as microscale,
mesoscale, and macroscale \citep{Sporns2005, Heuvel2017}. Microscale is in the
level of the neurons and synapses, macroscale is used to define brain regions
and large-scale communication pathways. Mesoscale is an intermediate level
between micro and macroscale, where connections between large portions of the
neuronal system are defined. A simple example of mesoscale network is the
mini-columns \citep{Stoop2013}. 

Neuronal networks are defined into structural or functional
\citep{Bullmore2011}. Functional networks are based on EEG, MEG or fMRI
measures \citep{Stam2012}. Functional networks of Alzheimer's patients present
increased path length when compared with healthy subjects \citep{Stam2007}. 

Structural connections can be characterized by diffusion weighted magnetic
resonance imaging (DW-MRI) and graph theory \citep{Lo2010}. DW-MRI analyzes
water diffusion in white matter, and together with fiber tractography it can be
used to identify structural connections in the brain \citep{Medina2007}. The
structural connection matrices of macaque and cats exhibit a complex structure
\citep{Hilgetag2000}. The presence of clusters and modular architecture in
structural connection matrices are observed by means of cortical thickness
measurements \citep{Chen2008}. Matrices with small-world properties and
exponentially truncated power law distribution were also reported
\citep{Gong2008}. 

In humans, the structural connection matrix mediates several complex cognitive
functions \citep{Bressler1995}. Abnormalities in structural networks were found
in patients with psychiatric disorders and neurodegenerative diseases
\citep{Stam2007,He2008,He2009,Yao2010,Pol2013,Stam2014}. Disconnection between
frontal and temporal cortices were observed in patients with Schizophrenia
\citep{Friston1995,Zalesky2011}. Hyperconnectivity in the frontal cortex were
reported in patients with Autism \citep{Courchesne2005}. Alzheimer's patient
showed increased path length and reduced global efficiency \citep{Lo2010}. The
alterations in brain networks are good indicators that network properties can
be used as biomarkers for clinical applications \citep{Kaiser2011}. 

Using diffusion tensor tractography, Lo et al. \citep{Lo2010} constructed
structural connection matrices of the human brain of healthy and Alzheimer's
subjects. The network is divided in 78 areas according to the automated
anatomic label template \citep{Tzourio-Mazoyer2002}. The connection between the
areas are defined in terms of the number of fibers, that were obtained through
fiber assignment by continuous tracking algorithm \citep{Mori1999}. 

In this work, we analyze the network properties of one structural connection
matrix related to a healthy subject and other related to a subject suffering of
specific neurodegenerative disease (Alzheimer). We demonstrate that similar
networks to these brain matrices can be constructed using Newman-Watts
procedure. 

In Section 2, we provide a brief discussion about the network representation of
the connectome. In Section 3, we introduce basic quantities that can be used to
quantify networks. In Section 4, we discuss the basic quantities of small-world
networks in the light of statistical analyses. In Section 5, we compare the
properties of human brain networks to small-world networks. In Section 6, we
present our final remarks. 


\section{Methodology}

\subsection{Properties of networks}

Networks properties provide information about segregation, integration and
influence \citep{Rubinov2010,Sporns2013}. Segregation properties are associated
with the presence of clusters or modules and integration properties are related
to the network ability to transmit information through its nodes. Segregation
and integration are linked with the network features while influence focus on
the node features proving information about the relevance of a node inside the
network. 

\subsubsection{Eigenvalues of the adjacency matrix}

The eigenvalues of the adjacency matrix $A$ are obtained by solving the
characteristic equation of $A$,
\begin{eqnarray}
det(A-\lambda I)=0,\label{eq1}
\end{eqnarray}
where $I$ is the identity matrix and the values of $\lambda$ that satisfy Eq.
(\ref{eq1}) are the eigenvalues \citep{Cvetkovic2008}. If the network is
symmetric, $A_{ij}=A_{ji}$, then all the eigenvalues are real.  

\subsubsection{Degree and node strength}

Degree $\kappa_i$ is the number of neighbors of a node $i$, 
\begin{eqnarray}
\kappa_i=\sum_{j=1}^{N}A_{ij},
\end{eqnarray}
where $N$ is the network size. It is considered one of the simplest measures to
provide information about the influence of the network. The degree distribution
is used to differentiate regular networks from random networks.

For weighted networks ($W_{ij}$), the use of node streng\-th $s_i$ instead of
degree $\kappa_i$ may be more appropriated \citep{Opsahl2010}. Node strength
$s_i$ is defined as the sum of the node connections, 
\begin{eqnarray}
s_i=\sum_{j=1}^{N}W_{ij}.
\end{eqnarray}

\subsubsection{Transitivity}

Transitivity $T$, Also known as clustering, is a measure of the segregation of
a network. The Transitivity $T$ is a measure of the amount of clustering
between the node $i$ and its $k_i$ neighbours, the maximum number of
connections between $i$ neighbors is $C_{\mbox{max}(i)}=k_i(k_i-1)/2$. $C_i$ is
defined as the ration between the number of active connections over the maximum
number of connections $C_{\mbox{max}(i)}$. The Transitivity $T$ is the average
over all nodes of the network.

The transitivity shows the effective proportion of the triangulation formed
between the sites as a measure of clustering capacity, $G(E,V)$. Then, $T$ is
calculated by the following proportional ratio
\begin{equation}
T=\frac{3 \delta(G)}{\tau(G)},
\end{equation}
where $\delta(G)$ or the number of triangles in graph G and $\tau(G)$ to denote
the number of triples in graph $G$ \citep{Schank2005}. 

One simple method is to use the arithmetic mean \citep{Opsahl2009}. If the
nodes $i$, $j$, and $k$ are connected, forming a triplet, the value of the
triplet is the arithmetic mean between $W_{ij}$ and $W_{jk}$. A triplet is
considered a close tripled when the nodes $i$, $j$, and $k$ are all connected
to each other. 

\subsubsection{Characteristic path length}

Characteristic path length $L$ measures the average of the shortest paths
$d_{ij}$ between all pairs of nodes in the network,
\begin{eqnarray}
L=\frac{2}{N (N-1)} \sum_{i=1}^{N}\sum_{j=1}^{N} d_{ij}.
\end{eqnarray}
This quantity is used for weighted and unweighted networks, it provides
information about the network integration. When dealing with diffusion process
and weigh\-ted networks, to calculate the shortest paths $d_{ij}$ the inverse of
the node strength should be used \citep{Opsahl2010}. For example, if
$W_{12}=2$, then $d_{ij}=1/2$, this approach considers that the higher is the node
strength the faster information can be diffused through it.

\subsubsection{Modularity}

Networks can be divided in two or more modules, the trivial solution is to
divide them into two modules, where one module has one node and another
module containing all the remaining nodes. Basically, the modular structure
is defined for any network and the question is to know the best method to
identify modules in complex networks. An optimized quantity to characterize the
modularity $Q$ was defined by Newman \citep{Newman2006}, that is given by
\begin{eqnarray}
Q=\frac{1}{4m} \sum_{ij} \left(A_{ij} -\frac{k_i k_j}{2m}\right)(s_i s_j + 1),
\end{eqnarray}
where $m=1/2 \sum_i k_i$, $s_i$, and $s_j$ are indices that depend on the group.
The network is divided in two groups, if the site $j$ belongs to group 1, then
$s_j=1$, if $j$ belongs to group 2, then $s_j=-1$. $Q$ can be either positive
or negative, positive values indicate the possible presence of community
structure. 

\subsubsection{Assortativity}

Assortativity $ASR$ is a measure of the tendency of high connected nodes to be
connected to others of similar degree $k$ \citep{Foster2010}. When high
connected nodes are more often connected to low connected nodes, the network
exhibits dissortative mixing. To define assortivity it is necessary to define
the remaining $q(k)$ and $p(k)$. The probability that a random node has a
degree $k$ is given by the degree distribution $p(k)$, however, the probability
to select a random edge is not proportional to $p(k)$ but to $kp(k)$, because
the most connected nodes receive more connections. Considering that node $i$ is
connected to node $j$ through a random selected edge, the remaining degree is
the number of nodes that leaves the node $j$, excluding node $i$. The
normalized remaining degree distribution is given by
\begin{eqnarray}
q(k)=\frac{(k+1)p(k)}{\sum_j j p(j)}.
\end{eqnarray}
The Assortativity ASR is defined as:
\begin{eqnarray}
\mbox{ASR}=\frac{1}{\sigma^2_q}\sum_{ij} ij(e(i,j)-q(i)q(j)),
\end{eqnarray}
where $\sigma^2_q$ is the variance of the remaining degree and $e(i,j)$ is the
joint probability distribution of the remaining degree of two nodes
\citep{Newman2002}. The Assortativity $A$ is defined in the interval
$-1 \leq A \leq 1$, when $A=1$ the network has perfect assortative mixed
patterns, $A=0$ indicates the network is not assortative and $A=-1$ means the
network is completely dissortative. 
 
\subsection{Statistical analysis}
 
\subsubsection{Generalized regression analysis}
 
A generalized linear model is made up of a combination of linear predictor with
link function, 
\begin{enumerate}
\item
Pedictor linear: $\eta_i=\beta_0+\beta_1 x_{1i}+\beta_1 x_{2i}\cdots
+\beta_p x_{pi}$
\item Link function: $g(\mu_i)=\eta$ for exponencial family of distributions
\[f(y,\theta,\phi)=\exp\left\{ \frac{y\theta-b(\theta)}{\phi-c(y,\phi)}
\right\},\] with
$$\left\{ \begin{array}{c} E(Y_i)=\mu_i=\frac{d[b(\theta_i)]}{d\theta}\\
Var(Y_i)=\phi^{-1}V_i\quad \phi^{-1}>0.\\
V=\frac{d\mu}{d\theta}\\
\end{array}\right.$$
\end{enumerate}

\subsubsection{Multivariate data analysis}

Multivariate analysis is a branch of statistics that deals with the
relationship between many variables, including the reduction of the number of
variables observed during an experiment. The main tools for multivariate data
analysis are principal component analysis (PCA) \citep{Gray2017}, factor
analysis \citep{Jhonson1998}, classifications \citep{Jhonson1998}, structural
equations models (SEM) \citep{Grace2016,Maruyama1998}, among other techniques.
In our case, the multivariate analysis of the data is useful to vary the
possible second order relationships between variables not directly correlated,
such as transitivity, assortativiness and the modularity of the human network.
At the end of this paper we will see how these measures are related using the
SEM \citep{Maruyama1998}.

\subsubsection{Development of a questionnaire}

We apply a questionnaire to a population in the small-world models artificially
generated with network size in $3$ to $100$ sites, from a single connection to
the global connection. The determination of sample \citep{Cochran1977}
\begin{eqnarray}
n=\frac{\sum_{N=3}^{100}\sum_{k=1}^{N} {{N}\choose{k}}\hat{p}\hat{q}z_{\alpha/2}^{2}}
{\hat{p}\hat{q}z_{\alpha/2}^{2}+(\sum_{N=3}^{100}\sum_{k=1}^{N} {{N}\choose{k}}-1)E^2},
\end{eqnarray}
for optimization $\hat{p}=\hat{q}=\frac{1}{2}$, with error $E\approx  \pm 3\%$
in $N=2499$ population with $z_{\alpha/2}$ is a z-score distribution with level
of significance $\alpha\approx 5\%$. In this case, the sample is $n\approx 492$
small-world models. This questionnaire is composed of $34$ variables or
questions about graph proprieties and applied to each small-world. Each model
randomly generated with a certain probability is measured with these
thirty-four variables.
 
We verify the quality of the questionnaire through Cronbach's alpha
\citep{Cronbach1951}
\begin{eqnarray}
\alpha=\frac{K}{K-1}\left(1-\frac{\sum_i \sigma^2_{Y_{i}}}{\sigma^2_{X_{i}}}\right),
\end{eqnarray}
where $K$ is the number of components, $\sigma^2_{X_{i}}$ is the variance of the
observed total test scores, and $\sigma^2_{Y_{i}}$ is the variance of the current
sample of generated small world. The questionnaire quality applied to
small-world networks is equal to 0.89, indicating a good Internal consistency
\citep{Cronbach1951,Maruyama1998}.
 
\begin{center}
\begin{table}[htpb]
\caption{Cronbach’s alpha.}
\begin{tabular}{r|l}
alpha & Internal consistency \\ \hline
$0.9 \leq \alpha $ & Excellent \\
$0.8 \leq \alpha \leq 0.9 $ & Good\\
$0.7 \leq \alpha \leq 0.8 $ & Acceptable \\
$0.6 \leq \alpha \leq 0.7 $ & Questionable\\
$0.5 \leq \alpha \leq 0.6 $ & Poor\\
$\alpha<0.5  $&  Unacceptable\\
\end{tabular}
\end{table}
 \end{center}
 
The questionnaire is composed of thirty-four questions (or variables) measured
directly in each artificial small-world network. Each variable measures an
important network property and the comparison between the human network and the
small-world model is given by means of these measures. The small global
templates are generated with $10$ sites up to $100$ sites. Each generated model
has different connections of its neighborhood between a single neighbor and the
global network. This way, $492$ samples of small-world models are produced. The
measure of sampling adequacy (MSA) through the Kaiser-Meyer-Olkin (KMO) test
indicates considered reasonable for value $KMO = 0.73$ \citep{KMO1974}. The KMO
and RMSA measures are given by
 \begin{equation}
KMO=\frac{\sum_{i=1}^k \sum_{j=1}^k r_{ij}}{\sum_{i=1}^k \sum_{j=1}^k r_{ij}^2
+a_{ij}^2}, 
\end{equation}
and
\begin{equation}
RMSA=\frac{\sum_{j=1}^k r_{ij}}{ \sum_{j=1}^k r_{ij}^2+a_{ij}^2},
\end{equation}
where $r_{ij}$ is the correlation matrix term and $a_{ij}$ is the
anti-image-correlation matrix term. In this method, the inverse correlation
matrix is close to the diagonal matrix. To verifies if matrix correlations is
statistical equivalent to an identity matrix, we use the Bartlett's test. The
basic hypothesis is that population's correlation matrix is an identity matrix
equivalent. In our variables group, p-value$<<0.05$ implies the rejection of
the null hypothesis and accepting the factorial analysis.    
 
\begin{table}[htpb!]
\centering
\caption{Variables of questionnaire - graph.}
\scriptsize
\label{questionnaire}
\begin{tabular}{r|l}
$N  :      $ & Number of nodes                        \\
$p  :      $ & Probability connection                 \\
$DE :      $ & Density edge                           \\
$L  :      $ & Average path length                    \\
$Rep:      $ & Reciprocity                            \\
$E:        $ & Edges                                  \\
$N:        $ & Vertices                               \\
$NL:       $ & number of links                        \\
$NL_{I}:    $ & Internal number of links               \\
$DL:       $ & link Density                           \\
$TST:      $ & Total System Throughput                \\ 
$TSFR:     $ & Total System Flow Rate                 \\
$Cn:       $ & Conectancie                            \\
$ALW:      $ & Average Link Weight                    \\
$ACT:      $ & Average Compartment Throughflow        \\
$Cp:       $ & Compartmentalization                   \\
$T:        $ & Transitivity                           \\
$ASR:      $ & Assortivity                            \\
$Ecc:      $ & Eccenticity                            \\
$D:        $ & Diameter                               \\
$I_{max}:   $ & Maximum interweaving                   \\
$I_{min}:   $ & Minimum interweaving                   \\
$Q:        $ & Modularity                             \\
$E:        $ & Eficience                              \\
$R:        $ & Radius                                 \\
$Kmax:     $ & Max K-Core                             \\
$Kmin:     $ & Min K-Core                             \\
$Kmean:    $ & Mean K-Core                            \\
$Iso:      $ & Isomorfism                             \\
$Auto:     $ & Automorphism                           \\
$EfcM:     $ & Number of edge with max efficiency     \\
$Efcm:     $ & Number of edge with minimal efficiency \\
$\lambda_p:$ & Principal eigenvalures                 \\
$det M:    $ & Determinant matrix 
\end{tabular}
\normalsize
\end{table}


\section{Results}

\subsection{Structural connection matrices}

\begin{figure}[h]
\centering
\includegraphics[width=0.9\linewidth,clip]{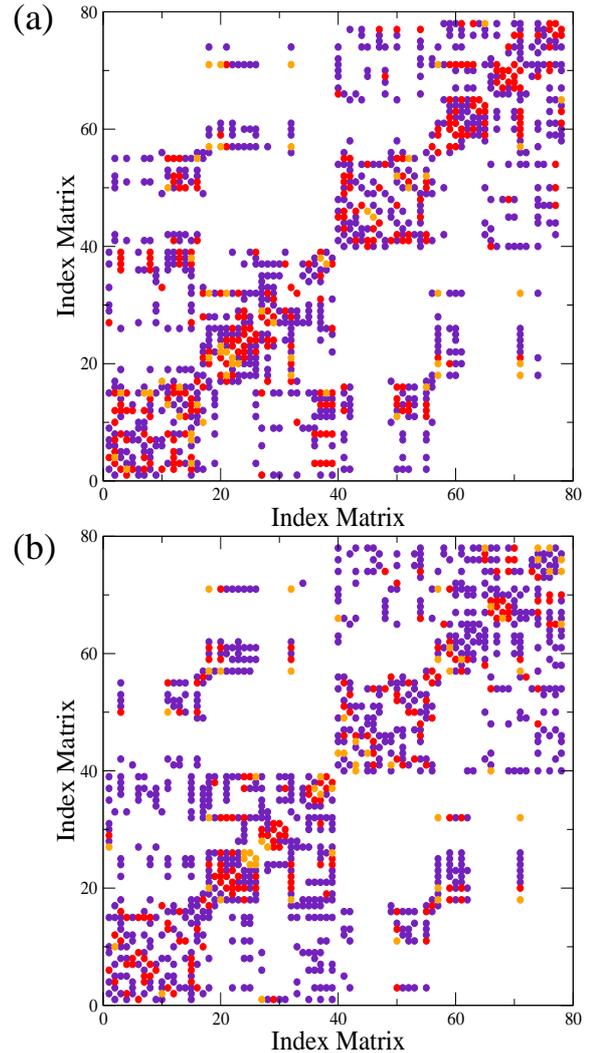}
\caption{Weighted connection matrix $W_{ij}$ for (a) healthy and (b) Alzheimer's
brains. The weight is associated with the intensity of connections: 0 (no
connections, white region), 1 (low density of connections, indigo circles), 2
(intermediate density of connections, red circles), and 3 (high density of
connections, orange circles).}
\label{fig1}
\end{figure}

\begin{figure}[h]
\centering
\includegraphics[width=1\linewidth,clip]{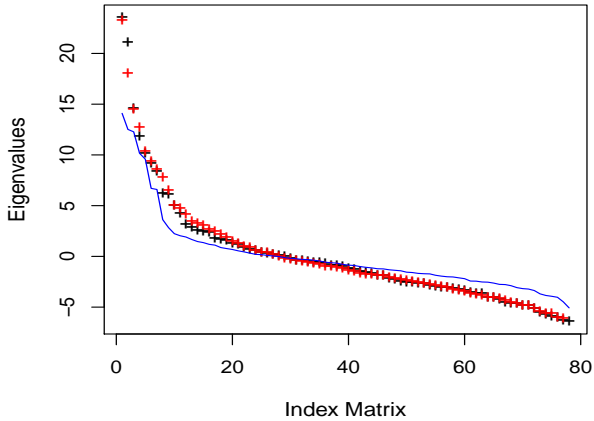}
\caption{Eigenvalue spectrum for the weighted matrices of Fig. \ref{fig1}.
Healthy structural connection matrix is in black and Alzheimer's structural
connection matrix is in red. Eigenvalue spectrum for small-world with $78$
nodes and connection probability is $p=0.0375$ in blue continue line.}
\label{fig2}
\end{figure}

In this work, we use two structural connection matrices, one for the healthy
brain (Fig. \ref{fig1}(a)) and other for the Alzheimer's brain (Fig.
\ref{fig1}(b)) \citep{Lo2010}. Both networks are weighted and symmetric, the
weight is associated with the intensity of connections and can assume five
values: 0 (no connections, white region), 1 (low density of connections, indigo
circles), 2 (intermediate density of connections, red circles), and 3 (high
density of connections, orange circles). The main results for the networks are
shown in Table \ref{table1}. Figure \ref{fig1} exhibits two adjacency matrix
connection $W_{ij}$: (a) healthy brain and (b) Alzheimer's brain. The
eigenvalues for these adjacency matrix are evaluated in Fig. \ref{fig2}. The
healthy structural connection matrix is in black and Alzheimer's structural
connection matrix is in red. The eigenvalue spectrum for small world with $78$
nodes is $p=0.0375$ (blue line). The ordinate eigenvalues are very close for
three structures matrix. The eigenvalues are equivalent when there is some
difference in the dispersion of adjacency matrix.

\begin{table}[htpb]
\centering
\caption{Network Indicators (W:Weighted Un:Unweighted).}
\label{table1}
\begin{tabular}{r|llll}
& \multicolumn{2}{c}{Healthy} & \multicolumn{2}{c}{Alzheimer}  \\  
                      & W          &  Un        &  W        & Un     \\ \hline
Transitivity $T$      & 0.578      & 0.578      & 0.560     & 0.559  \\
Assortivity  $ASR$    & 0.081      & 0.010      & 0.226     & 0.125  \\ 
Path Length $L$       & 2.248      & 2.248      & 2.281     & 2.281  \\
Modularity $M_1$      & 0.451      & 0.423      & 0.483     & 0.428  \\
$\sigma$              & 23.27      & 0.590      & 7.534     & 0.590  \\
Average Degree        & 8.000      & 1.383      & 17.487    & 1.383  \\
\end{tabular}
\end{table}

\subsection{Small-world networks}

A network with small-world properties can be generated by means of different
methods \citep{Newman2000}. The most common method was developed by Watts and
Strogatz \citep{Newman2000}, where the regular edges are replaced by random
edges. When about 1$\%$ of the total edges are replaced, the network exhibits
high transitivity and low path length \citep{Watts1998}. For our analysis, we
consider an alternative method, where instead of randomly replace regular edges
by random edges, we only add random edges, known as Newman-Watts procedure
\citep{Newman2003}. We add $pNK$ new random edges, where $N$ is the network
size, $K$ is the regular network degree, and $p$ is the probability to add new
edges. We vary $p$ and identify the small-world properties comparing $T$ and
$L$ with the values of the regular network $T(0)$ and $L(0)$. We find
\begin{itemize}
\item
Healthy brain
\begin{equation}
DL:\mbox{link density}\longrightarrow\frac{DL}{2}=\frac{13.3333}{2}\approx 7, 
\end{equation}
\item
Alzheimer's brain
\begin{equation}
DL:\mbox{link density}\longrightarrow\frac{DL}{2}=\frac{13.38462}{2}\approx 7. 
\end{equation}
\end{itemize}

The number of nodes in the human graph is $78$ sites and the average degree is
$7$ neighbors per node. Due to this fact we create the equivalent connection
network under these conditions to depend exclusively on the probability of
calling. It is generated small-world graph with number of nodes $N=78$, $7$
neighbours per node, and without the likelihood of connection ($p=0$), implying
in average length $L_0=3.27273$ and transitivity $C_0=0.69231$. For healthy
human structural connection matrix, we find $L_{\mbox{healthy human}}=2.24875$
and $C_{\mbox{healthy human}}=0.57813$. Then, we obtain 
\begin{eqnarray}
\frac{C_{\mbox{healthy human}}}{C_0} & = & 0.8350739, \\ \nonumber
\frac{L_{\mbox{healthy human}}}{L_0} & = & 0.6871175, \\ \nonumber 
             & \Rightarrow & \frac{C}{C_0}>\frac{L}{L_0}, \\ \nonumber
\frac{C_{\mbox{Alzheimer's structural}}}{C_0} & = & 0.8087265, \\ \nonumber
\frac{L_{\mbox{Alzheimer's structural}}}{L_0} & = & 0.6971917, \\ \nonumber 
             &\Rightarrow& \frac{C}{C_0}>\frac{L}{L_0}. \nonumber
\end{eqnarray}

$C/C_0$ is approximately $1.4\%$ more than Alzheimer's disease human cluster
standardized, and $L/L_0$ is approximately $3.1\%$ less than Alzheimer's
disease human. The propagated information in Alzheimer's disease human
presents greater difficulty for diffusion of information in network, becoming
more complex than healthy human matrix. Therefore, there seems to be a
relationship between the transfer of the network and its grouping, i.e.,
relations between assortivity, modularity and transitivity.
 
In correlation matrix, we present a statistical correlations $r_{ij}$ for
variables assortivity (ASR), modularity (Q), and transitivity (T),
respectively, for small-world samples classes used in RMSA and KMO analysis.
All values are low and indicate the lack of direct correlation. 
  
\subsubsection{Regression analysis}
 
In a convenience sample, for $n=429$ small-world type networks, five
replicas are executed to create variation within the others. The dispersion of
the transitivity according to the logarithm of the connection probability shows
a decay adjusted by generalized model Gaussian family with link identity 
\begin{eqnarray}
f(y|\mu,\sigma^2) & = & \exp\left[\frac{1}{\sigma^2}\left(y\mu-\frac{\mu^2}{2}
\right)+\left(-\frac{1}{2}\ln(2\pi\sigma^2) \right.\right.\nonumber \\
& & \left.\left. -\frac{y^2}{2\sigma^2}\right)\right],
\nonumber
\end{eqnarray}
resulting in the following regression
\begin{equation}
T_{SW}= 0.300314 -0.029338 \ln p.
\label{SWW33}
\end{equation}
For instance, when the probability connection is $p=0.00091188196$, then
$\log p$ is equal to $-7$. The result is approximately $0.50568$. This
probability value $p$ is in agreement with the probability of small-world
connection. When the connection probability increases, the dispersion of the
transitivity value increases as well. On the other hand, the decrease in
probability linkage causes the dispersion to become smaller and more
concentrated, characterizing a good small-world region. $T_H=T_M=T_{SW}$ is
valid for the small-world model. Another feature of Eq. \ref{SWW33} is its rate
of transitivity in relation to the log of the probability,
\begin{equation}
\frac{d T_{SW}}{d \ln p}=-0.029338 \approx -3\%.
\label{SWW23}
\end{equation}
The ratio of transitivity to $\ln p$ is equal to the loss value in the
small-world model when we compare the matrix of healthy human adjacency with
disease Alzheimer human matrix.

The transfer rate and the rate of flow in the network with Alzheimer's exhibit
a drop equal to that caused in the transitivity when compared with the human
network in the normal state, as shown in Table \ref{tabelaTL}. It suggests that
the rate of transfer and rate of flow for people with Alzheimer's disease
declines with $5.4\%$, possibly due to the fall in transitivity in $3.1\%$.
   
\begin{table}[htpb]
\centering
\caption{Transitivity loss (weighted).}
\label{tabelaTL}
\begin{tabular}{r|ccc}
               &  Healthy  & Alzheimer   & loss     \\\hline
Transitivity   & 0.57813   & 0.5598876   & $3.1\%$  \\ 
Nodes          & 78        & 78          &          \\
Links total    & 1040      & 1044        &          \\
Transfer rate  & 1438      & 1364        & $5.4\%$  \\
Leak rate      & 1438      & 1364        & $5.4\%$  \\
\end{tabular}
\end{table}

\subsubsection{Assortativity}

One of the most difficult measure to be statistically analyzed is the
assortivity of the network. Due to the fact that the network topology of
small-world is very sensitive to the probability of (re)connection. This can be
verified in healthy and Alzheimer's human matrices. for $T=0.57$, the
assortivity shows very different values. However, the assortiveness is $2.5$
times higher for the Alzheimer's brain than for healthy (Table \ref{table1}).

The Alzheimer's brain is more assortive than the healthy brain, this means that
in the Alzheimer network the nodes with high degree are, in average, connected
with other nodes of high degree more intensely than the healthy human network.
We calculate the assortativity distribution for small-world networks for
$n=492$ samples. In a sample, for example the small-world $N=10$ and second
order connection, the assortiveness presents an sample average of
$<ASR>=-0.01335938$ not being statistically zero according to t-Student test
for the hypothesis $H_0:\mu_{ASR}=0$. There is no significant evidence to
support the null hypothesis for a p-value $<<0.05$, inclining us to accept the
hypothesis that assortiveness in the sample question is, in fact, negative.
The network is on average disassortative. This does not imply the formation of
positive assortiveness as verified in the graph.

\subsubsection{Probabilistic valley}

The probabilistic valley is a region where the small-world structure behaves by
sequences of abrupt changes. It is precisely in this region that we identify
abrupt behaviors of assortativiness, given the equivalent modularity and
transitivity. These three measures of the small-world model that are equivalent
to the measurements of the human matrix are found in this valley. The
probabilistic voucher is developed through the structural equations model
(SEM), in which it is related indirectly to assortivity, transitivity and
modularity. As assortiveness represents the equivalent of the correlation
between the links of the sites of a network, we write the assortivity in
function of the transitivity and the modularity of the network for determinate
probabilistic valley.
 
The probabilistic valley region indicates a possible existence of probability
as a function of the modularity, transitivity and efficiency, that it is in
agreement with the SEM analysis. This indicates that there is a possible
dependence on the functions of modularity, assortiveness, transitivity and
efficiency, according to the SEM analysis. In the same region random overflow
occurs in assortivity increasing transitivity and modularity. The increase in
assortivity and transitivity implies in the decay of the connection
probability, which confirms that the probability value decreases. A more
detailed view of the level curve with the modularity in the abscissa of the
assortative at the ordinate reveals a complex structure of the curves. Outside
this region the value of the assortiveness is zero or close to zero.
 
\begin{figure}[h]
\centering
\includegraphics[width=0.85\linewidth,clip]{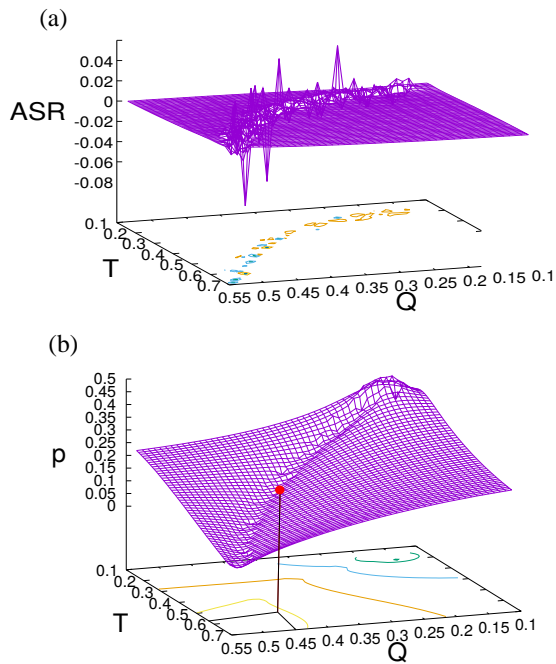}
\caption{Probabilistic valley as functions of modularity and transitivity for: (a) Assortivity and (b) probability of non-local connections.}
\label{fig3}
\end{figure}

The valley has many interesting behavior. The assortivity, transitivity and
modularity measures exist on\-ly because there are valley probabilistic. To
generate Fig. \ref{fig3}, we consider a set of $100$ independent models of
small-world networks starting with $10$ sites up to the amount $100$ sites. In
all models, we vary the probability of linkage between the non-coupling state
($p=10^{-6}$) and the overall state ($p=1$). The red dot in Fig. \ref{fig3} is
located in the region where the modularity ($\approx 0.45$) and transitivity
($\approx 0.58$) have values equal to the results found through the human
matrix of healthy individuals. The same point coincides with the result of the
assortiveness ($\approx -0.02$) in the Alzheimer's human matrix. This graph
located in this point have $78$ sites with $7$ connections neighbors and
probability range $8.10^{-8}<p<0.5$. 
  
\section{Discussion}

\begin{figure}[h]
\centering
\includegraphics[width=0.8\linewidth,clip]{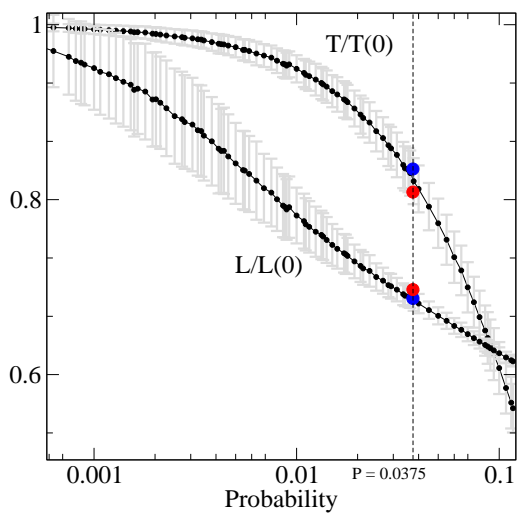}
\caption{$L/L(0)$ and $T/T(0)$ as a function of the probability. The results for
healthy and Alzheimer's structural connection matrix are shown by blue and red
balls, respectively.}
\label{fig4}
\end{figure}

The human networks are located in the probabilistic valley. We locate the
healthy and Alzheimer's human networks within the small-world samples. The
superposition both red and blue dots in evolutionary average length measures
and transitivity graphics in connection probability is $p=0.0375$, as shown by
the vertical dotted line of Fig. \ref{fig4}. This ordered pair 
$\left(\frac{T}{T(0)},\frac{L}{L(0)}\right)=(0.8350739,0.6871175)$ for $p=0.0375$
is exhibited by the blue dots in Fig. \ref{fig4}. The same technique is used to
find the ordered pair representing Alzheimer's structural connection matrix for
\begin{equation}
\left(\frac{T}{T(0)},\frac{L}{L(0)}\right)=(0.8087265 ,0.6971917).
\end{equation}
In this case, it is represented by the red dots in same figure. 

In fact, when we select the region of the transitivity of Fig. \ref{fig4} in
the value of the probability of connection, we have that the difference between
the blue and red points is 3.1\%, which represents the healthy individual and
Alzheimer's disease, respectively. The human Alzheimer's network exhibits
greater difficulty in the transmission of information due to the fall of
transitivity in the network. On the other hand, Fig. \ref{fig4} also shows that
the path length ($L$) is larger than the case of healthy individuals. The
individuals with Alzheimer's disease have a decrease in the effectiveness of
the transitivity.

\begin{figure}[h]
\centering
\includegraphics[width=0.8\linewidth,clip]{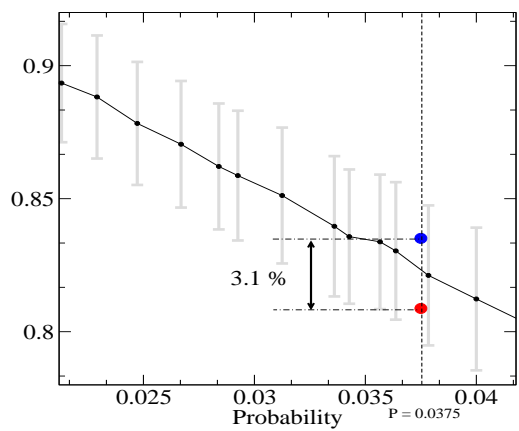}
\caption{Magnification of $T/T(0)$ as a function of the probability of Fig. \ref{fig4}.}
\label{fig5}
\end{figure}

The weighted connection matrix eigenvalues $W_{ij}$ is useful for the comparison
between the matrix structures. In Fig. \ref{fig2}, we display the eigenvalue
spectrum for both networks of Fig. \ref{fig4}. We verify that the eigenvalues
of both networks are similar. The eigenvalues of the proposed small-world model
are very close to the human adjacency matrices. The approximation of the
variation of the transitivity region in Fig. \ref{fig4} can be seen in Fig.
\ref{fig5}. This figure show us two points, one blue and another red that
represent the health human and Alzheimer's disease, respectively. The difference
between the the healthy and Alzheimer's brains is about $3.1\%$.  

Table \ref{tabelaH} shows that the average path length value of the healthy
brain is very close to small-world network. The transitivity , assortivity,
eccentricity, and modularity are almost identical. In fact, the relationship
between a human graph and a small-world structure is pertinent. In Table
\ref{tabelaA}, we see that the Alzheimers's brain and small-world network have
similar values, except the assortivity value, $\varepsilon=59.62\%$.

\begin{table}[h]
\centering
\caption{Values of the healthy brain and small-world network.}
\label{tabelaH}
\scriptsize
\begin{tabular}{r|ccc}
                    & Health human & Small-world  & error            \\
                    &    (real)    &  (Simulated) & ($\varepsilon$)  \\ \hline
Average path length & 2.2487       & 2.1964       & +2.32\%          \\
Density of links    & 13.333       & 14.000       & -5.00\%          \\
Transitivity        & 0.5781       & 0.5386       & +6.79\%          \\
Assortivity         & 0.0815       & 0.0882       & -8.32\%          \\
Eccentricity        & 3.6667       & 3.5128       & +4.20\%          \\
Modularity          & 0.4515       & 0.4889       & -8.27\%          \\ 
\end{tabular}
\end{table}
\normalsize

\begin{table}[h]
\centering
\caption{Values of the Alzheimer's brain and small-world network.}
\label{tabelaA}
\scriptsize
\begin{tabular}{r|ccc}
                    & Alzheimer's         & Small-world  & error           \\
                    & brain (real)        &  (Simulated) & ($\varepsilon$) \\
\hline
Average path length & 2.28172             & 2.1964       & +3.74\%         \\
Density of links    & 13.3846             & 14.000       & -4.59\%         \\
Transitivity        & 0.55989             & 0.5386       & +3.80\%         \\
Assortivity         & 0.21846             & 0.0882       & +59.62\%        \\
Eccentricity        & 3.76923             & 3.5128       & +6.80\%         \\
Modularity          & 0.49083             & 0.4889       & +0.39\%         \\
\end{tabular}
\end{table}
\normalsize

\section{Conclusions}

In this work, we show that small-world networks can be used to mimic brain
networks. Comparing the healthy human matrices and the small-world model with
probability of connection about $3.75\%$. It is evident the proximity of the
measures indicators of the graphs, such as the average path length that
presented $97\%$ of proximity with the result of the human matrix.

We find a relation of construction among the variables associated with the
transmission of information in the network, such as the transitivity (0.57813
for the healthy brain and 0.5386 for the small-world network), the assortivity
(0.08151 forthe healthy brain and 0.08829 for the small-world network), the
eccentricity (3.66667 for the healthy brain and 3.51282 for the small-world
network), and the modularity (0.45157 for the healthy brain and 0.48891 for the
small-world network), whose values are very close to each other. In all four
measures, we obtained errors ($\varepsilon$) smaller than 10\% in the
measurements up or down. This characteristic was verified in the variation of
the transitivity and the length of the average path as a function of the
probability. In both cases, the healthy human network and the human network
for the Alzheimer's brain were within the simulated region for a small-world
sample, thus indicating a close linkage probability of $3.75\%$. Exactly a
small-world model in this region, for an equivalent assertiveness value, have
very similar graph properties thus demonstrating that the human network
(diseased or not) behave as a small-world network.

We verify that the healthy brain can be mimicked by networks with small-world
properties. The network indicators of the Alzheimer's brain are almost
identical with the small-world network, except the assortivity. Therefore, the
assortivity could be a diagnostic tool to identify Alzheimer's brain.

\begin{acknowledgements}
We wish to thank the Brazilian government agencies: Funda\c c\~ao Arauc\'aria,
CNPq (420699/2018-0, 407543/2018-0), FAPESP (2015/50122-0, 2018/03211-6), and
CAPES for partial financial support.
\end{acknowledgements}

\bibliographystyle{elsarticle-harv}

\end{document}